\documentclass[twocolumn,showpacs,superscriptaddress,amssymb,10pt]{revtex4-1}

\usepackage{graphicx}
\usepackage{dcolumn}
\usepackage{bm}
\usepackage{booktabs}
\usepackage{enumitem}
\usepackage{longtable}
\usepackage{adjustbox}
\usepackage{graphicx} 
\usepackage{amsmath}   
\usepackage{color}

\newcommand{\be}{\begin{eqnarray}}
\newcommand{\ee}{\end{eqnarray}}

\newcommand{\balpha}{\bm{\alpha}}

\newcommand{\bmu}{\bm{\mu}}

\newcommand{\bfJ}{{\bf J}}
\newcommand{\bfL}{{\bf L}}
\newcommand{\bfS}{{\bf S}}
\newcommand{\bfT}{{\bf T}}
\newcommand{\rmd}{{\rm d}}

\newcommand{\bfp}{{\bf p}}
\newcommand{\bfr}{{\bf r}}

\newcommand{\mub}{\mu_0}

\newcommand{\bfTnr}{\bfT_{\rm nr}}

\newcommand{\jji}{J_{i}}

\begin{document}

\preprint{}
\title{
Line strengths of QED-sensitive forbidden transitions 
in B-, Al-, F- and Cl-like ions }

\author{M.~Bilal}
\affiliation{Helmholtz-Institut Jena, Fr\"o{}belstieg 3, D-07743 Jena, Germany}%
\affiliation{Theoretisch-Physikalisches Institut, Friedrich-Schiller-Universit\"at Jena, Max-Wien-Platz 1, D-07743
Jena, Germany}%

\author{A.~V.~Volotka}
\affiliation{Helmholtz-Institut Jena, Fr\"o{}belstieg 3, D-07743 Jena, Germany}%

\author{R. Beerwerth}
\affiliation{Helmholtz-Institut Jena, Fr\"o{}belstieg 3, D-07743 Jena, Germany}%
\affiliation{Theoretisch-Physikalisches Institut, Friedrich-Schiller-Universit\"at Jena, Max-Wien-Platz 1, D-07743
Jena, Germany}%

\author{S.~Fritzsche}
\affiliation{Helmholtz-Institut Jena, Fr\"o{}belstieg 3, D-07743 Jena, Germany}%
\affiliation{Theoretisch-Physikalisches Institut, Friedrich-Schiller-Universit\"at Jena, Max-Wien-Platz 1, D-07743
Jena, Germany}


\begin{abstract}

The magnetic dipole (M1) line strength between the fine-structure levels of the ground configurations in 
B-, F-, Al- and Cl-like ions are calculated for the four elements argon, iron, molybdenum and tungsten. 
Systematically enlarged multiconfiguration Dirac-Hartree-Fock~(MCDHF) wave functions are 
employed to account for the interelectronic interaction with the Breit interaction included in first-order 
perturbation theory. The QED corrections are evaluated to all orders in $\alpha Z$ utilizing an effective
potential approach. The calculated line strengths are compared with the results of other theories. The M1 
transition rates are reported using accurate energies from the literature. Moreover, the lifetimes in the
range of millisecond to picoseconds are predicted including the contributions also from the transition rate 
due to the E2 transition channel. The discrepancies of the predicted rates from those available from the 
literature are discussed and a benchmark dataset of theoretical lifetimes is provided to support future 
experiments.  

\end{abstract}


\maketitle

\section{Introduction}
\label{Sec.Introduction}

Transition energies and transition rates are two fundamental properties of atomic states. 
Therefore, a detailed analysis and comparison of theoretical predictions with experimental 
observations may provide crucial insight into our basic understanding of the atomic structure. 
For level energies, there exists a number of cases where very high accuracy has been achieved 
from both theory and experiment, and helped to make quantum electrodynamic (QED) and many-body relativistic 
effects visible. For example, QED has been tested at the level of 7.2\% for the M1 
transition energy between fine-structure levels of the ground configuration in 
B-like Ar \cite{Draganic03,Artemyev07,Mackel11,Artemyev13}. For the transition rates and line 
strengths, in contrast, the accuracy level is often not yet sufficient to test QED and many body 
relativistic effects. This is partially due to theory and partially due to experiment. We 
know that transition rates depend on higher power of transition energy and non-diagonal matrix
elements of the multipolar electromagnetic operators. In contrast to transition 
energies, there is no variational principle available that defines a minimum condition for the 
optimization of non-diagonal matrix elements. For this reason, the many-body relativistic effects 
are more difficult to capture. While the experimental accuracy is typically 2\% and higher due
to systematic and statistical errors \cite{Serpa98,Moehs98,Moehs99,Trabert00,Beier03,Smith05}, 
this has been found insufficient to explore relativistic and QED effects, see for details 
the reviews \cite{Trabert08,Trabert14}. However, there are two remarkable exceptions where an accuracy 
of the order of 0.1\% is claimed by efficiently controlling the systematic and statistical errors. 
Both of these lifetime measurements were performed at the Heidelberg electron beam ion trap~(HD-EBIT).
The measured lifetime is reported as 9.573(4)(5)(stat/syst) ms for the $ 2s^2 2p$ $^2P_{3/2}$ level in
B-like Ar \cite{Lapierre05} and 16.726(+20/-10) ms for the $ 3s^2 3p $ $^2P_{3/2}$ level in 
Al-like Fe \cite{Brenner2007}. Both of these levels decay dominantly via a magnetic dipole (M1) transition
between the fine-structure levels of the ground configuration. 

In the case of an M1 transition between the fine-structure levels of the same configuration the non-diagonal
matrix element i.e., the line strength, is less sensitive than for E1 allowed transitions. This is 
because in the nonrelativistic limit, the M1 line strength is insensitive to the description of the 
many-electron wave functions. In other words, (almost) all correlation corrections are of relativistic 
origin and, therefore, suppressed by a factor $\alpha Z$ ($Z$ is the nuclear charge). For such transitions the 
line strengths are especially sensitive to the QED contributions. For instance, the leading QED effect of 
an order $\alpha$, so called electron anomalous magnetic moment (EAMM) correction, contributes to 
0.46 \%  \cite{Tupit05}. Therefore, such M1 transition rates can be calculated very precisely and may be
used as a benchmark for comparison with the experiment. 

During recent years, various \textit{ab initio} calculations have been reported for the M1 line strength  
between the $ 2s^2 2p$ $ ^2P_{3/2} \,-\, ^2P_{1/2} $ levels in B-like ions \cite{Tupit05,volotka06,Volotka08,Rynkun12,Marq12,Fisch16} and 
$ 3s^2 3p $ $ ^2P_{3/2} \,-\, ^2P_{1/2} $ in Al-like ions \cite{Vilkas03,Sant09}. In particular, the line strength 
of the $ 2s^2 2p$ $ ^2P_{3/2} \,-\, ^2P_{1/2} $ transition in B-like Ar has been evaluated with a relative
uncertainty of only $10^{-5}$ \cite{Tupit05}. However, all these calculated line strengths combined
with experimental transition energies tend to predict shorter lifetimes than measured 
experimentally \cite{Lapierre05,Brenner2007}. In fact, the deviation between theory and experiment is of the 
order of the EAMM, which led to a speculation about the correctness of the inclusion of the EAMM into the transition amplitude. 
Let us note here that such high-precision measurements are available only from the HD-EBIT in the millisecond range. 
In the future, however, precise experiments
for various lifetime and transition energy domains and by different techniques will 
hopefully solve the present discrepancy. Therefore, there is strong need for a theoretical analysis of these
systems where relativistic correlations and the QED contributions can be quantified as a benchmark principal
for these experiments. 
 
We here present a detailed study for the line strengths of QED-sensitive forbidden transitions 
between the fine-structure levels of the ground configurations in B-, Al-, F- and Cl-like ions. 
The ground configurations of B-like and Al-like ions have a valence $p$-electron in the L shell and M shell, respectively. 
These configurations are also quite similar to the ground configurations of F-like and Cl-like ions but with a $p$-shell vacancy
in the L shell and M shell, respectively. The major difference between the two systems of ions is the flip of fine-structure levels 
where the excited $^2P$ ground-state levels dominantly decay through M1 transition. We will combine our accurate line strengths with 
accurate experimental or theoretical transition energies and predict lifetimes in the millisecond to picoseconds range.

The rest of the paper is structured as follows. 
A short description of the underlying theory and our calculations are described in Sec. \ref{theory}. In Sec. \ref{results},
we present a detailed comparison of our calculated line strengths with other theories. Here we add contributions from the 
E2 channel and the M1 channel. From the total transition rates we predict the lifetimes and compare
with available experiments. Finally, our main findings are summarized
in Sec. \ref{con}. Atomic units ($\hbar = m = e = 1$) are used throughout the paper.  

\section{Theory and calculations}\label{theory}
\subsection{Theory - Basic Formulas }
The magnetic dipole transition probability from an upper state $i$ to a lower state $f$ is expressed in 
terms of the line strength as 
\begin{equation}\label{trans}
{\rm W} = \frac4{3}\frac{\omega^3}{c^3}\mub^2 \frac {\rm S}{2\jji+1} , 
\end{equation}
where $\mub$ denotes the Bohr magneton, $c$ is the speed of light, $ \omega = E_i - E_f$
the transition energy and where the line strength is, 
\begin{equation}\label{srel}
{\rm S} = \frac{18c^4}{\omega^2}\left|  \langle \Psi{_f}
\parallel \bfT
  \parallel \Psi{_i} \rangle  \right|^2 \,.
\end{equation}
$\bfT$ is the M1 transition operator given as
\begin{eqnarray}\label{trans-opr} 
  \bfT = \frac{1}{\sqrt{2}} \, j_1(\omega r/c) \,
  \frac{[ \balpha \times \bfr ]}{r} \, = \, 
  \frac{\sqrt2}{r} j_1(\omega r/c) \bmu \,.
\end{eqnarray}
Here $\bmu =  -[ \bfr  \times \balpha ]/2$ is relativistic magnetic moment operator, 
$\balpha$ is Dirac matrix and $j_1$ is the spherical Bessel function. 
 
In the nonrelativistic limit the expansion of $j_1(\omega r/c)$ can be restricted to the first term, 
and this gives rise to the more familiar M1 transition operator
\begin{equation}
  \bfTnr \,=\, - \frac{\sqrt{2}}{3} \, \frac{\omega}{c}  \mub \, (\bfL + 2 \bfS) \,.
\end{equation}
where $\bfL$ and $\bfS$ are the orbital and spin angular momentum
operators, respectively. In the $LS$-coupling scheme, which is realized in the nonrelativistic case,
the M1 line strength is nonzero only between
fine-structure levels with $\Delta J = \pm 1$.
The reduced matrix element of $\bfTnr$ within the $LS$-coupling
is given by

\begin{equation} \nonumber
\begin{split}
  \langle J_f \parallel \bfTnr \parallel J_i \rangle & = - \,
  \frac{\sqrt{2}}{3} \, \frac{\omega}{c}\mub \langle J_f \parallel
  (\bfJ + \bfS) \parallel J_i \rangle \\
            & =  - \, \frac{\sqrt{2}}{3}
  \frac{\omega}{c}\mub \langle J_f \parallel \bfS \parallel J_i \rangle \,,
\end{split}
\end{equation}

which implies,
\begin{equation}\label{snr}
{\rm S_{nr}} = \left|  \langle J{_f}
\parallel \bfS
  \parallel J{_i} \rangle  \right|^2 \,.
\end{equation}
Therefore, in the nonrelativistic limit the line strength ${\rm S_{nr}}$ is 
completely determined by the quantum numbers of the initial and final states and does not depend on the radial part of the 
many-electron wave functions of the initial and final states. 
For the $ ^2P_{1/2} \,-\, ^2P_{3/2}$ and 
$ ^2P_{3/2} \,-\, ^2P_{1/2} $, fine-structure transitions, the nonrelativistic line strength results in the value of 4/3. 

The total line strength can be calculated by adding different corrections to the nonrelativistic line strength as follows,
\begin{equation}\label{tot-S}
\mathrm{S} = \mathrm{S_{\rm nr} + \Delta S_{\rm D} + \Delta S_{\rm CI,C} + 
\Delta S_{\rm CI,B} + \Delta S_{\rm QED} + \Delta S_{\rm rec}}\\
\end{equation}

Here $\Delta {\rm S_D}$ is the correction due to the \textit{relativistic} motion of the electrons as described by the 
(single-electron) Dirac equation. This corrections is calculated as a difference between line strength 
evaluated between Eq. (\ref{snr}) and Eq. (\ref{srel}). In Eq. (\ref{srel})
the initial and final state wave functions are linear combination of Slater determinants constructed in terms 
of one-electron Dirac wave functions which are the solution of the non-interacting one electron Dirac Hamiltonian,
\begin{equation} 
{\hat{h}_{\rm D}} = c\balpha\cdot \boldsymbol p + (\beta - 1)c^{2}-V(r)
\end{equation}
where $ V $ is the potential of a two parameter Fermi nuclear charge distribution, 
$\beta$ is  Dirac-matrix and $ c $ is the speed of light in atomic units.

The next two terms in Eq. (\ref{tot-S}) are due to the relativistic interelectronic interaction (correlations). While the first 
term $\Delta {\rm S_{CI,C}}$ arises from the Coulomb interaction, and the second $\Delta {\rm S_{CI,B}}$ 
occurs due to the Breit interaction. Both of these terms are evaluated in details in Sec. \ref{CI}. 
The next correction $\Delta {\rm S_{ QED}}$ originates from QED diagrams, namely, the self-energy diagrams. 
It is calculated here to all orders in $\alpha Z$. The evaluation of this term is described in Sec. \ref{qed}.

Finally, $\Delta {\rm S_{rec}}$ is the correction to the line strength due to the finite nuclear mass effect. 
This effect can be calculated only by using a rigorous 
QED approach as described in Ref. \cite{Volotka08}. According to this approach the
recoil corrected magnetic moment operator is given by
\begin{equation}
  \bmu = -\mu_0\,\Bigl(\bfL + 2\bfS - \frac{1}{M}\sum_{i,j}[\bfr_i\times\bfp_j]\Bigr)\,,
\end{equation}
where $M$ is mass of the nucleus. Hence the correction to the line strength due to the nuclear recoil 
can be written as 
\begin{equation}
\begin{split}
  \Delta {\rm S_{rec}} & \simeq
     - 2 \langle J_f \parallel
  (\bfL + 2\bfS) \parallel J_i \rangle \\
 & \times \langle J_f \parallel 
  \frac{1}{M}\sum_{i,j}[\bfr_i \times \bfp_j] \parallel J_i \rangle \,.
\end{split}
\end{equation} 
However, these contributions are
very small at the present level of accuracy compared to the leading
nonrelativistic value 4/3. For example  $\Delta {\rm S_{rec}}$ 
amounts to 0.000021, 0.000015, 0.000009 and 0.000005 for B-like Ar, Fe, Mo, and W ions, respectively.
Therefore, we do not present these values in our final table of the various contribution to line strengths.  

\subsection{Interelectronic-interaction corrections}\label{CI}

In order to evaluate the interelectronic correlation correction arising due to Coulomb interaction $\Delta {\rm S_{CI,C}}$,
we apply systematically enlarged many-electrons wave functions by using the latest version of 
the general purpose relativistic atomic structure package \textsc{Grasp2K} \cite{Jonson13}. This package implements the
multiconfiguration Dirac-Hartree-Fock~(MCDHF) method in $jj$-coupling \cite{Grant07}. In this method, $\Psi$ is an 
atomic state function $\Psi(\Gamma; \pi J)$ for a state label $\Gamma $, where $ J $ is the total angular momentum quantum 
number and $ \pi $ is the parity. It is approximated by a linear combination of configuration state 
functions~(CSFs) of the same symmetry:
\begin{eqnarray}
\Psi{(\Gamma ; \pi J)}  = \displaystyle \sum_{j=1}^{n_{c}}c_{j}\Phi(\gamma_{j}; \pi J),
\end{eqnarray}
where $n_{c}$ is the number of CSFs, $ c_{j} $ are the mixing coefficients and $ \gamma_{j} $ denotes the
orbital occupancy and angular coupling scheme of the $j$-th CSF. The configuration state functions 
$\Phi(\gamma_{j}; \pi J)$ are a linear combination of Slater determinants of one electron Dirac spinors, 
\begin{eqnarray}
\Phi(r) = \frac{1}{r}
\begin{pmatrix}
P_{nk}\left(r\right)\chi_{\kappa}^{m}(\theta,\varphi) \\
iQ_{nk}\left(r\right)\chi_{-\kappa}^{m}(\theta,\varphi)
\end{pmatrix}.\label{eq:2}
\end{eqnarray}
Here, $ \kappa $ is relativistic angular momentum quantum number, $P_{nk}\left(r\right)$ and $Q_{nk}\left(r\right)$ are 
the large and small radial components of the one electron wave functions represented on a logarithmic grid, and 
$\chi_{\kappa}^{m}$ is the spinor spherical harmonic. The radial part of the Dirac 
orbitals and the expansion coefficients $ c_{j} $ are optimized to self consistency from a set of equations 
which results from applying the variational principle on a weighted energy functional of the states in 
Dirac-Coulomb approximation \cite{Dyall} where the Dirac-Coulomb Hamiltonian $ \hat{H}_{DC}$ is, 
\begin{eqnarray}\nonumber
\hat{H}_{\rm DC}=\sum_{i=1}^{N}\bigg[c\balpha_{i}\cdot \boldsymbol{p}_{i}+(\beta_{i}-1)c^{2}-V(r_{i})\bigg]+
\sum_{i<j}^{N}\frac{1}{r_{ij}}.\\
\end{eqnarray}

We first performed the calculations for the lowest-order approximation. For this 
the wave functions for the 
states with $J = 1/2 $ and $ J = 3/2 $ are calculated within the basis of the multi-reference (MR) configurations.
The CSFs in the MR set are generated from the configurations \{$1s^2 2s^2 2p$,  $1s^2 2p^3$\}, 
\{$1s^2 2s^2 2p^6 3s^2 3p$, $1s^2 2s^2 2p^6 3p^3$\}, 
$1s^2 2s^2 2p^5$ and $1s^2 2s^2 2p^6 3s^2 3p^5$ for the B-, Al-, F- and Cl-like ions, respectively.  
After the initial calculations, the wave functions are systematically improved by performing 
MCDHF calculations for each new layer of correlation orbitals and keeping the previous calculated orbitals fixed. For each new
layer of correlation orbitals the basis of CSFs is expanded by including further single (S) and double (D) virtual
excitations from the configurations defining the MR set to the active set of orbitals. The active set of orbitals is 
spanned by the orbitals with a principal quantum number $n \leq 7$ and with azimuthal quantum number $\l \leq 6$. 

Following each of the MCDHF calculations, separate relativistic configuration interaction (RCI) calculations are performed to further 
improve the initial and final state wave functions. These allowed us to evaluate the correction due to the Breit 
interaction $\Delta {\rm S_{CI,B}}$ to the line strength. For these calculations the Dirac-Coulomb Breit Hamiltonian 
${\hat{H}_{\rm DCB} = \hat{H}_{\rm DC} + \hat{H}_{\rm Breit}}$ is used where 
%
%
%
\begin{equation}
{\hat{H}_{\rm Breit}}  =  -\sum_{i<j}^{N} \frac{1}{2r_{ij}} \Bigg [ \balpha_{i}\cdot \balpha_{j} 
+ \frac{(\balpha_{i} \cdot {\bm {r}_{ij}})(\balpha_{j} \cdot {\bm {r}_{ij})}}{r_{ij}^2}\Bigg ].
\end{equation}
%

The sum of these two corrections gives rise to the total relativistic interelectronic-interaction correction $\Delta{\rm S_{CI}}$.
These contributions are presented in Table \ref{tab:S_B} as a function of the size
of the increasing active set labeled by the highest principal quantum number $n$ of the of orbitals considered
for the correlations. For the sake of brevity, we present the results only for the Fe ions.  
As seen from Table \ref{tab:S_B}, the convergence with regard to the size of the 
active set is fairly achieved which allows us to set an absolute
uncertainty for the interelectronic-interaction correction to a range
$1 \times 10^{-5}$ - $5 \times 10^{-5}$ depending on the particular ion.
\begin{table}[b] 
\caption{ Interelectronic-interaction correction $\Delta{\rm S_{CI}}$ to the M1 line strength
of the transition between the fine-structure levels of the ground configuration for B-, F-, Al- and 
Cl-like Fe ions. The MCDHF and RCI methods are employed to evaluate these corrections
considering Coulomb and Breit type interactions.
They are presented as a function of the 
size of the increasing active set~(AS) labeled by the highest principal 
quantum number $n$ of the orbitals starting from the MR set.}\label{tab:S_B} 

\begin{ruledtabular}
\begin{tabular}{ccccc}
AS   &  B-like Fe & F-like Fe & Al-like Fe & Cl-like Fe \\
\colrule
MR   &     0.00053  &   0.00154  &    0.00149  &    0.00167   \\ 
3    &     0.00052  &   0.00146  &    0.00148  &    0.00165   \\ 
4    &     0.00049  &   0.00151  &    0.00146  &    0.00164   \\ 
5    &     0.00050  &   0.00153  &    0.00145  &    0.00164   \\ 
6    &     0.00045  &   0.00146  &    0.00147  &    0.00165   \\ 
7    &     0.00045  &   0.00145  &    0.00147  &    0.00164   \\ 
                                                                 
\end{tabular}
\end{ruledtabular}
\end{table}

\subsection{QED correction}\label{qed}

The QED correction to the M1 line strength $\Delta {\rm S_{QED}}$ can be derived
in lowest order in $\alpha Z$ by modifying the M1 transition operator of the
atomic magnetic moment for the EAMM, as discussed in details in Ref.~\cite{Tupit05}.
The contribution of the EAMM amounts to $\Delta {\rm S_{QED,EAMM}} = 0.00618$. In
Ref.~\cite{volotka06}, moreover, the one-loop QED correction was calculated for several
B-like ions to all orders in $\alpha Z$ within the so-called original Furry picture
- and by taking into account only the Coulomb potential of the nucleus. We now consider 
an \textit{extended} Furry picture which includes a local screening potential in the
unperturbed Hamiltonian, and extend the calculations for F-, Al-, and Cl-like systems.
This extension enables us to account partially for the screening QED
corrections by evaluating only one-electron QED diagrams. In the extended Furry
picture, we solve the Dirac equation with an effective spherically symmetric
potential treating the interaction with the external Coulomb potential of the
nucleus and the local screening potential exact to all orders. We employ here the
core-Hartree screening potential, which is given by the expression
\begin{eqnarray}
  V_{\rm scr}(r) = \int_0^\infty \rmd r'\frac{1}{r_>}\rho_{\rm core}(r')\,.
\end{eqnarray}
Here $\rho_{\rm core}$ denotes the total radial charge density distribution of
the core electrons
\begin{eqnarray}
  \rho_{\rm core}(r) = \sum_c \left[ P_c^2(r) + Q_c^2(r) \right]\,,\nonumber\\
  \int_0^\infty dr \rho_{\rm core}(r) = n_c\,,
\end{eqnarray}
where $n_c$ is the number of the core electrons, i.e., $n_c = 4,\,8,\,12,\,16$ for
B-, F-, Al-, and Cl-like ions, respectively. This screening potential is generated
self-consistently by solving the Dirac equation until the energies of the core and
valence states become stable with the relative accuracy of $10^{-9}$. To estimate 
the sensitivity of the result on the choice of the potential
several tests have been performed with other screening potentials: Kohn-Sham,
Dirac-Hartree, and Dirac-Slater constructed for the initial as well as for
the final state. It has been found out, that a relative difference between
results obtained with different potentials does not exceed $5 \times 10^{-4}$.
Overall, therefore, the uncertainty is dominated by a numerical error, which
is everywhere smaller than $10^{-5}$.

\begin{figure}
\includegraphics[width=0.4\textwidth]{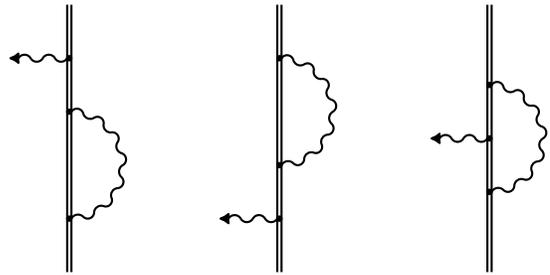}
\caption {Feynman diagrams that represent the self-energy correction to the line
strength. The wavy line indicates the photon propagator and the double line
indicates the bound-electron wave functions and propagators in the effective
potential being the sum of Coulomb and screening potentials. The single photon
emission is depicted by the wavy line with arrow.}
\label{fig:se}
\end{figure}

The one-loop QED correction to the line strength consists of the self-energy and
vacuum-polarization terms. However, the vacuum-polarization correction previously evaluated in
the Uehling approximation appears to be two $-$ four orders of magnitude
smaller than the self-energy correction beyond the EAMM approximation \cite{volotka06D}. For this reason,
we neglect the vacuum-polarization term in the present consideration. The self-energy
contribution is given by the diagrams depicted in Fig.~\ref{fig:se}.
While the formulas derived in Ref.~\cite{volotka06} in the original Furry picture remain
formally the same, let us remind that the Dirac spectrum is now generated by
solving the Dirac equation with the effective potential. We make use of the implementation of
Ref.~\cite{volotka06} where a detailed description of
these calculations for B-like ions is presented.

\section{results and discussion}\label{results}

Table \ref{tab:S1} lists different corrections to the nonrelativistic line strength for the M1 transition
between the fine-structure levels of the ground configurations in B-like and F-like Ar, Fe, Mo and W ions. 
These corrections refer to two different systems of $p$ subshells, one with a single valence electron and other with a
single vacancy in the L shell. Table \ref{tab:S2} shows similar results as Table \ref{tab:S1} 
but for the M shell for Al-like and Cl-like Fe, Mo and W ions.
As seen from Tables \ref{tab:S1} and \ref{tab:S2} the relativistic correction $\Delta {\rm S_D}$ is most important.
Its value increases by an order of magnitude for the Mo and W ions when compared to Ar and Fe ions.  
The interelectronic-interaction correction arising due to Breit interaction turns out to be relatively small 
as compared to corrections arising due to Coulomb interaction. For Mo and W ions these corrections provide 
essential contributions to the total line strengths. The next important correction arises from the 
self-energy (QED). Generally, the lowest order QED correction i.e., the inclusion of EAMM to the transition operator, 
is considered enough for such type of transitions. As discussed in sections \ref{qed} this 
correction amounts to 0.00618 the total line strength. However, the present estimates of the QED correction 
show that the inclusion of the EAMM is enough only for low $Z$. For the heavier systems the present rigorous calculations 
of QED corrections within an extended Furry picture approach are necessary. 
Finally, all corrections sum to the total line strength S. In order to
estimate the total uncertainty we have to note here that the contribution of
the negative-energy excitations is not taken into account in present
calculations. Since the value of the negative-continuum term strongly
depends on an employed one-electron basis functions \cite{Tupit05} varying
them we estimate its contribution to be less than half of the correlation
effect. The uncertainties of individual terms presented in Tables \ref{tab:S1} and
\ref{tab:S2} as discussed above in corresponding subsections are much smaller than
the uncertainty due to the negative-continuum contribution. Therefore, the
total uncertainty of the line strength S obtained is fully determined by
missing negative-continuum contribution.

In order to compare our results with previous computations, we make use of the line strength from M1 transition rates
based on Eq. (\ref{trans}). We took the same transition energies that have been used in respective calculations. 
It should be noted that we have added the QED contribution to the line strengths in respective theoretical values 
wherever this effect has not been considered. 
For B-like Ar, which has received much attention over the last decade, our result of 
total line strength 1.33714(28) corroborates the two exceptionally agreeing calculations
from Tupitsyn \textit{et al.} \cite{Tupit05} and from Froese Fisher \textit{et al.} \cite{Fisch16}.
Tupitsyn \textit{et al.} \cite{Tupit05} reported the line strength value 1.33693(26) for B-like Ar. 
Additionally, Froese Fisher \textit{et al.} \cite{Fisch16} reported a line strength 1.3372 for B-like Ar. 
Froese Fisher \textit{et al.} \cite{Fisch16} further extended their calculations for the B-like isoelectronic sequence
until $Z = 42$. As seen from Table \ref{tab:S1} our results also agree with B-like Fe and B-like Mo.
Our results of the line strengths for all other systems under present study are in agreement 
with other available theories. The only exception is with the relativistic 
coupled-cluster calculations for which the line strength of 1.356(5) in F-like Ar of Ref. \cite{Nandy16} is overestimated
and the line strength of 1.3270 in F-like Fe of Ref. \cite{Nandy14} is underestimated. 
This may be due to the incorrect handling of intruder states in the implementation of coupled-cluster theory in 
open shell systems \cite{Derev12}. Overall, our calculations have reached an accuracy of $10^{-4} - 10^{-5}$ 
for the QED sensitive M1 line strengths between the fine structure levels of the same configuration. 
As a result, the present calculations provide a theoretical prerequisite for a test of QED effects in the 
line strengths of various ions. 

In Table \ref{lifetime} we present the lifetimes $\tau_{\rm pres}$ (in seconds) calculated for 
the $(2s^2 2p)\ ^2P_{3/2}$ level in B-like ions, the $(2s^2 2p^5)\ ^2P_{1/2}$ level 
in F-like ions, the $(3s^2 3p)\ ^2P_{3/2}$ level in Al-like ions as well as for the $(3s^2 3p^5)\ ^2P_{1/2}$ 
level in Cl-like ions. Here, $\tau_{\rm exp}$ are experimental lifetimes and respective experimental uncertainties are 
given in parentheses. ${\rm A_ {M1}}$ is the present transition rate from the M1 channel. 
We used the best available transition energies from the literature for the calculation of the transition rate. For B-like ions we
applied the transition energies from the rigorous QED treatment of Artemyev \textit{et al.} \cite{Artemyev07,Artemyev13}, and for 
the rest of the ions we used transition energies from the NIST database \cite{NIST}. Here ${\rm A_{E2}}$ is 
the transition rate from the E2 channel in length form. For the E2 transition rate, we make use of the same wave function
expansion as for the M1 transition rate, in addition to that the length and the velocity gauges of the E2 
line strength were in good agreement. Let us note that the 
present uncertainties in transition rates and lifetimes are due to uncertainties in the calculation of the M1 line strengths only. 
They are given within the parentheses. For the present level of accuracy, the uncertainties due to the E2 transition channel are very small.
However, the uncertainties in the transition energies will increase total uncertainties in our calculations accordingly. 

As seen from Table \ref{lifetime}, our predicted lifetimes for B-like Ar and for Al-like Fe disagree with both experiments
at the HD-EBIT \cite{Lapierre05,Brenner2007}. In contrast, the comparison of our 
predicted lifetime for F-like Ar with the experiment at the LLNL-EBIT \cite{Trabert00} and 
for the lifetime of Cl-like Fe with the experiment at the HD-EBIT \cite{Brenner09} shows very good agreement. For Cl-like Fe, our lifetime
also agrees well with the extrapolated lifetime of Ref. \cite{Trabert04} which is resulted in an experimental study along Cl-like Co,
Ni and Cu ions. These experiments with an uncertainty larger than 0.5\% are however not sensitive enough to test 
the underlying relativistic correlations and the leading QED effects.  
New experiments with the soft x-ray free electron laser (FLASH) and a new EBIT \cite{Epp07} along with the pump 
probe x-ray laser experiments \cite{Trabert14a} are hopeful to provide experimental data for the transitions with
short lifetimes in so far inaccessible energy ranges. We believe that our calculations will support such future experiments
for transitions with different frequencies and lifetimes.

\section{conclusion}\label{con}
 
In this paper, we have presented highly accurate calculations for the line
strengths of QED-sensitive forbidden transitions by utilizing the multiconfiguration
Dirac-Hartree-Fock and relativistic configuration interaction methods.
We have extended
the high precision evaluations previously performed for the middle $Z$ B-like ions \cite{Tupit05,Fisch16}
to higher $Z$ as well as to different systems such as F-, Al-, and Cl-like ions.
In our systematically enlarged wave functions, we incorporated all important electron correlations and the effects of
relativity by taking the Coulomb and Breit interactions into account.
The obtained line strengths are further improved by rigorous calculations of the QED correction within an extended
Furry picture approach. We used up-to-date accurate transition energies for the calculations
of the M1 transition rates and reported lifetimes in the millisecond to picoseconds range.
We believe that our accurate theoretical predictions provide the prerequisite for a test of QED by lifetime measurements
at different frequencies and timescales.
This will help to find a reason for the present discrepancies between theory and experiment for B-like Ar
and Al-like Fe. Apart from testing atomic structure theory, such experiments in the future agreeing with the theoretical
investigations will be very helpful for terrestrial and astrophysical plasma diagnostics.

\begin{table*} 
\caption{ Individual corrections to the M1 nonrelativistic line strength {\rm S}$_{\rm nr}$ = 4/3
for the $(2s^2 2p) \, ^2P_{1/2} \,-\, ^2P_{3/2}$ transition  
in B-like as well as for the $(2s^2 2p^5)\ ^2P_{3/2} \,-\, ^2P_{1/2}$ transitions in F-like
Ar, Fe, Mo and W ions. The total line strength (${\rm S}$) is compared with other theories.
The uncertainties involved in the calculation of line strengths are given
within the parentheses.}\label{tab:S1}
\begin{ruledtabular}
\begin{tabular}{ldddd}

 & \multicolumn{1}{c} {\rm Ar} & \multicolumn{1}{c} {\rm Fe} & \multicolumn{1}{c} {\rm Mo} & \multicolumn{1}{c} {\rm W} \\
    
\colrule
B-like                                    &                                      &                          &                                   &                                                      \\
                                          &                                      &                          &                                   &                                                      \\
$\Delta {\rm S_D}$                        &   -0.00295                           &     -0.00633             &    -0.01800                       &    -0.07402                                          \\
$\Delta {\rm S_{CI,C}}$                   &   0.00056                            &     0.00038              &    -0.00247                       &    -0.00530                                          \\
$\Delta {\rm S_{CI,B}}$                   &   0.00001                            &     0.00007              &    0.00042                        &    0.00166                                           \\
$\Delta {\rm S_{QED}}$                    &   0.00617                            &     0.00615              &    0.00606                        &    0.00567                                           \\

${\rm S}$                                 &   1.33714(28)                        &     1.33362(23)           &    1.3194(10)                     &    1.2614(18)                                           \\
                                          &   1.3372\footnotemark[1]             &     1.3337\footnotemark[1]&    1.3197\footnotemark[1]         &                                                      \\
                                          &   1.33693(26)\footnotemark[2]        &     1.333\footnotemark[3]$*$ &                                   &                                                      \\
                                          &   1.337\footnotemark[3]$*$              &     1.333\footnotemark[4]$*$ &                                   &                                                      \\
                                          &   1.337\footnotemark[4]$*$              &                          &                                    &                                                      \\
                                          &                                      &                          &                                    &                                                      \\
\colrule
F-like                                    &                                      &                          &                                   &                                                      \\
                                          &                                      &                          &                                   &                                                      \\
$\Delta {\rm S_D}$                        &     -0.00295                         &     -0.00633             &    -0.01800                       &     -0.07402                                         \\
$\Delta {\rm S_{CI,C}}$                   &     0.00094                          &     0.00143              &    0.00258                        &     0.00569                                          \\
$\Delta {\rm S_{CI,B}}$                   &     0.00000                          &     0.00002              &    0.00006                        &     0.00036                                          \\
$\Delta {\rm S_{QED}}$                    &     0.00617                          &     0.00615              &    0.00607                        &     0.00568                                          \\

${\rm S}$                                 &     1.33749(47)                      &     1.33460(70)          &    1.3240(13)                     &     1.2710(30)                                        \\
                                          &     1.356(5)\footnotemark[5]$*$         &     1.335\footnotemark[7]$*$&    1.324\footnotemark[8]$*$          &     1.271\footnotemark[6] $*$                            \\
                                          &     1.338\footnotemark[6]$*$            &     1.334\footnotemark[6]$*$&    1.324\footnotemark[6]$*$          &     1.271\footnotemark[9]$*$                            \\
                                          &                                      &     1.3270\footnotemark[10]$*$&  1.3211\footnotemark[10]$*$        &                                                      \\

\end{tabular}
\end{ruledtabular}
\begin{tabular}{p{25cm}} \\[0.1cm]
\footnotemark[1]{MCDHF theory by Froese Fischer \textit{et al.} \cite{Fisch16}.}                \\
\footnotemark[2]{MC-DFS theory by  Tupitsyn \textit{et al.} \cite{Tupit05}.}                     \\
\footnotemark[3]{MCDF theory by Rynkun \textit{et al.} \cite{Rynkun12}.}                       \\
\footnotemark[4]{MCDF theory by Marques \textit{et al.} \cite{Marq12}.}                         \\
\footnotemark[5]{Relativistic coupled-cluster theory by Nandy \cite{Nandy16}.}                    \\
\footnotemark[6]{MCDF theory  by J\"onnson \textit{et al.} \cite{Jonnson13}.}                   \\
\footnotemark[7]{MCDF theory by Jonauskas \textit{et al.} \cite{Jonaus04}.}                     \\
\footnotemark[8]{MCDF theory by Aggarwal and Keenan \cite{Aggar16}.}                            \\
\footnotemark[9]{MCDF theory by Aggarwal and Keenan \cite{Aggar16b}.}                            \\
\footnotemark[10]{Relatively coupled-cluster theory by Nandy and Sahoo \cite{Nandy14}.}          \\                               
* The original values are corrected by adding the QED correction obtained here.  
\end{tabular}
\end{table*}                                          
\begin{table*} 
\caption{ Individual corrections to the M1 nonrelativistic line strength ${\rm S_{nr}}$ = 4/3
for the $(3s^2 3p)\, ^2P_{1/2} \,-\, ^2P_{3/2}$ transition in Al-like as well as for the
$(3s^2 3p^5)\ ^2P_{3/2} \,-\, ^2P_{1/2}$ transition in Cl-like Fe, Mo and W ions. 
The total line strength (${\rm S}$) is compared with other theories.
The uncertainties involved in the calculation of line strengths are given
within the parentheses.}\label{tab:S2}

\begin{ruledtabular}
\begin{tabular}{lddd}
          
 &  \multicolumn{1}{c} {\rm Fe} & \multicolumn{1}{c} {\rm Mo} & \multicolumn{1}{c} {\rm W} \\
    
\colrule

Al-like                                    &                          &                                   &                                                      \\
                                           &                          &                                   &                                                      \\
$\Delta {\rm S_D}$                         &     -0.00302             &    -0.00950                       &     -0.05025                                         \\
$\Delta {\rm S_{CI,C}}$                    &     0.00146              &    0.00230                        &     0.00340                                          \\
$\Delta {\rm S_{CI,B}}$                    &     0.00001              &    0.00007                        &     0.00054                                          \\
$\Delta {\rm S_{QED}}$                     &     0.00617              &    0.00614                        &     0.00595                                          \\

${\rm S}$                                  &     1.33797(73)          &    1.3324(13)                     &     1.2928(20)                                       \\
                                           &     1.336\footnotemark[1]$*$ &                                  &                                                      \\
                                           &     1.337\footnotemark[2] &                                  &                                                      \\
                                           &                          &                                   &                                                      \\
\colrule                                                                                                                                                    
                                           &                          &                                   &                                                      \\
Cl-like                                    &                          &                                   &                                                      \\
                                           &                          &                                   &                                                      \\
$\Delta {\rm S_D}$                         &     -0.00302             &    -0.00950                       &     -0.05025                                         \\
$\Delta {\rm S_{CI,C}}$                    &     0.00164              &    0.00294                        &     0.00615                                          \\
$\Delta {\rm S_{CI,B}}$                    &     0.00000              &    0.00005                        &     0.00029                                          \\
$\Delta {\rm S_{QED}}$                     &     0.00617              &    0.00614                        &     0.00595                                          \\

${\rm S}$                                  &     1.3381(18)           &    1.3330(15)                     &     1.2955(32)                                          \\
                                           &     1.338\footnotemark[3]$*$&                                   &     1.295\footnotemark[4]$*$                                                \\
                                           &     1.338\footnotemark[5]&                                   &     1.29\footnotemark[6]$*$                                                 \\

\end{tabular}
\end{ruledtabular}
\begin{tabular}{p{25cm}} \\[0.1cm]
\footnotemark[1]{MR-MP theory by Vilkas and Ishikawa \cite{Vilkas03}.}       \\
\footnotemark[2]{MR-MP theory by Santana \textit{et al.} \cite{Sant09}.}                  \\
\footnotemark[3]{B-spline single-particle orbitals method by Moehs \textit{et al.} \cite{Moehs01}.}                                \\
\footnotemark[4]{MCDF method by Aggarwal and Keenan \cite{Aggar14}.}        \\
\footnotemark[5]{MR-MP theory by Ishikawa \textit{et al.} \cite{Ishikawa10}.}         \\
\footnotemark[6]{MCDF theory by Singh and Puri \cite{Gajen16}.}             \\
* The original values are corrected by adding the QED correction obtained here.  
\end{tabular}
\end{table*}

\begin{table*} 
\caption{Lifetimes $\tau_{\rm pres}$ (in seconds) calculated for the 
$(2s^2 2p)\ ^2P_{3/2}$ level in B-like ions, the $(2s^2 2p^5)\ ^2P_{1/2}$ level 
in F-like ions, the $(3s^2 3p)\ ^2P_{3/2}$ level in Al-like ions and the $(3s^2 3p^5)\ ^2P_{1/2}$ 
level in Cl-like ions compared with experimental lifetimes ($\tau_{\rm exp}$).  
${\rm A_ {M1}}$ is the present transition rate (in $s^{-1}$) from the M1 channel and $A_{\rm E2}$ is 
the transition rate (in $s^{-1}$) from the E2 channel. The values of the transition energy used for the 
present lifetime calculations are given in cm$^{-1}$ and corresponding transition 
wavelengths $\lambda$ in \AA. The uncertainties involved in the calculation of transition 
rate and lifetime arising due to uncertainties in the line strengths are given within the 
parentheses. The numbers given in the square brackets denote powers of 10.}\label{lifetime}
\begin{ruledtabular}
\begin{tabular}{lllcccr}                                                
Ions         &      Energy     &        $\lambda$  & $A_{\rm M1}$ &$A_{\rm E2}$& $\tau_{\rm pres}$&
$\tau_{\rm exp}$\\
\colrule
B-like        &                 &                  &                     &                    &                     &                                                    \\
              &                 &                  &                     &                    &                     &                                                    \\
Ar$^{13+}$    &    22656.92     &       4413.663   &    1.0487(02)[+02]  &      1.86[-03]     &      9.5354(20)[-03]&   9.573(4)(5)[-03]\footnotemark[1]                    \\
              &                 &                  &                     &                    &                     &   8.7(5)[-03]\footnotemark[2]                      \\
              &                 &                  &                     &                    &                     &   9.12(18)[-03]\footnotemark[3]                    \\
              &                 &                  &                     &                    &                     &   9.70(15)[-03]\footnotemark[4]                    \\
              &                 &                  &                     &                    &                     &                                                    \\
Fe$^{21+}$    &    118310.243   &       845.235    &    1.4893(03)[+04]  &      1.37[+00]     &      6.7141(11)[-05]&                                                    \\
Mo$^{37+}$    &    964437.459   &       103.687    &    7.9810(60)[+06]  &      6.00[+03]     &      1.2520(09)[-07]&                                                    \\
W$^{69+}$     &    11802649.713 &       8.473      &    1.3985(20)[+10]  &      1.25[+08]     &      7.0874(10)[-11]&                                                    \\
              &                 &                  &                     &                    &                     &                                                    \\
Al-like       &                 &                  &                     &                    &                     &                                                    \\
              &                 &                  &                     &                    &                     &                                                    \\
Fe$^{13+}$    &    18852.5      &       5304.336   &    6.0455(33)[+01]  &      1.49[-02]     &      1.6537(09)[-02]&      1.6726(+20/-10)[-02]\footnotemark[5]          \\
              &                 &                  &                     &                    &                     &      1.7(2)[-02]\footnotemark[6]                   \\
              &                 &                  &                     &                    &                     &      1.674(12)[-02]\footnotemark[7]                \\
              &                 &                  &                     &                    &                     &      1.752(29)[-02]\footnotemark[8]                \\
              &                 &                  &                     &                    &                     &                                                    \\
Mo$^{29+}$    &    204020       &       490.148    &    7.6299(74)[+04]  &      1.67[+02]     &     1.3078(12)[-05] &                                                    \\
W$^{61+}$     &    2933400      &       34.090     &    2.2008(34)[+08]  &      6.30[+06]     &     4.4174(70)[-09] &                                                    \\
              &                 &                  &                     &                    &                     &                                                    \\
F-like        &                 &                  &                     &                    &                     &                                                    \\
              &                 &                  &                     &                    &                     &                                                    \\
Ar$^{9+}$     &    18067.494    &       5534.802   &    1.0639(04)[+02]  &      2.11[-03]     &     9.3994(33)[-03] &     9.32(12)[-03]\footnotemark[4]                  \\
Fe$^{17+}$    &    102579       &       974.858    &    1.9428(10)[+04]  &      1.94[+00]     &     5.1466(26)[-05] &                                                    \\
Mo$^{33+}$    &    886305       &       112.828    &    1.2432(12)[+07]  &      9.77[+03]     &     8.0372(79)[-08] &                                                    \\
W$^{65+}$     &    11202000     &       8.927      &    2.4097(57)[+10]  &      2.16[+08]     &     4.1131(98)[-11] &                                                    \\
              &                 &                  &                     &                    &                     &                                                    \\
Cl-like       &                 &                  &                     &                    &                     &                                                    \\
              &                 &                  &                     &                    &                     &                                                    \\
              &                 &                  &                     &                    &                     &                                                    \\
Fe$^{9+}$     &    15683.14     &       6376.274   &    6.9615(93)[+01]  &      1.52[-02]     &     1.4362(19)[-02] &   1.42(2)[-02]\footnotemark[9]                     \\
              &                 &                  &                     &                    &                     &   1.441(14)[-02]\footnotemark[10]                  \\
              &                 &                  &                     &                    &                     &   1.364(25)[-02]\footnotemark[8]                   \\
              &                 &                  &                     &                    &                     &                                                    \\
Mo$^{25+}$    &    186950       &       534.902    &    1.1746(13)[+05]  &      2.41[+02]     &     8.4959(96)[-06] &                                                    \\
W$^{57+}$     &    2796000      &       35.765     &    3.8190(94)[+08]  &      1.05[+07]     &     2.5485(65)[-09] &                                                    \\
              &                 &                  &                     &                    &                     &                                                    \\
                                                                                              
\end{tabular}
\end{ruledtabular}
\begin{tabular}{p{25cm}} \\[0.1cm]
\footnotemark[1]{HD-EBIT experiment by Lapierre \textit{et al.}. \cite{Lapierre05}.}\\
\footnotemark[2]{NIST-EBIT experiment by Serpa \textit{et al.}\cite{Serpa98}.}\\
\footnotemark[3]{ECRIS in a Kingdon ion trap experiment by Moehs \textit{et al.} \cite{Moehs98}.}\\
\footnotemark[4]{LLNL-EBIT by Tr\"abert \textit{et al.} \cite{Trabert00}.}\\
\footnotemark[5]{HD-EBIT experiment by Brenner \textit{et al.} \cite{Brenner2007}.}\\
\footnotemark[6]{ECRIS in a Kingdon ion trap experiment by Smith \textit{et al.} \cite{Smith05}.}\\
\footnotemark[7]{LLNL-EBIT experiment by Beiersdorfer \textit{et al.} \cite{Beier03}.}\\
\footnotemark[8]{ECRIS in a Kingdon ion trap by Moehs and Church \cite{Moehs99}.}\\
\footnotemark[9]{HD-EBIT experiment by Brenner \textit{et al.} \cite{Brenner09}.}\\
\footnotemark[10]{TSR measurements at the Max Planck Institute for Nuclear Physics, Heidelberg, Germany by Tr\"abert \textit{et al.} \cite{Trabert04}.}\\
\end{tabular}
\end{table*}

\clearpage
\begin{acknowledgments}
Discussions with I. I. Tupitsyn are gratefully acknowledged.  
\end{acknowledgments}


\begin{thebibliography}{99}                                                                                                                                                        

\bibitem{Draganic03}
I. Dragani\'c, J. R. Crespo L\'opez-Urrutia, R. DuBois, S. Fritzsche, 
V. M. Shabaev, R. Soria Orts, I. I. Tupitsyn, Y. Zou, and J. Ullrich,
Phys. Rev. Lett. {\bf 91}, 183001 (2003).

\bibitem{Artemyev07} A. N. Artemyev, V. M. Shabaev, I. I. Tupitsyn, 
G. Plunien, and V. A. Yerokhin, 
Phys. Rev. Lett. {\bf 98}, 173004 (2007).

\bibitem{Mackel11} V. M\"ackel, R. Klawitter, G. Brenner,
J. R. Crespo L\'opez-Urrutia, and J. Ullrich,  
Phys. Rev. Lett. {\bf 107}, 143002 (2011).

\bibitem{Artemyev13}
A. N. Artemyev, V. M. Shabaev, I. I. Tupitsyn, G. Plunien, 
A. Surzhykov, and S. Fritzsche,
Phys. Rev. A {\bf 88}, 032518 (2013).

\bibitem{Moehs98}
D. P. Moehs and D. A. Church, 
Phys. Rev. A {\bf 58}, 1111 (1998).

\bibitem{Serpa98}
F. G. Serpa, J. D. Gillaspy, and E. Tr\"abert,
J. Phys. B {\bf 31}, 3345 (1998).

\bibitem{Moehs99}
D. P. Moehs and D. A. Church,
Astrophys. J. {\bf 516}, L111 (1999).

\bibitem{Trabert00}
E. Tr\"abert, P. Beiersdorfer, S. B. Utter,
G. V. Brown, H. Chen, C. L. Harris, P. A. Neill,
D. W. Savin, and A. J. Smith, Astrophys. J. {\bf 541}, 506 (2000).

\bibitem{Beier03}
P. Beiersdorfer, E. Tr\"abert, and E. H. Pinnington,
Astrophys. J. {\bf 587}, 836 (2003).

\bibitem{Smith05}
S. J. Smith, A. Chutjian, and J. A. Lozano, 
Phys. Rev. A {\bf 72}, 062504 (2005).
͒
\bibitem{Trabert08}
E. Tr\"abert,
Can. J. Phys., {\bf 86}, 73 (2008).

\bibitem{Trabert14}
E. Tr\"abert,
Atoms, {\bf 2}, 15 (2014).

\bibitem{Lapierre05}
A. Lapierre, U. D. Jentschura,
J. R. Crespo L\'opez-Urrutia, J. Braun, G. Brenner,
H. Bruhns, D. Fischer, A. J. Gonz\'alez Mart\'inez, Z. Harman,
W. R. Johnson, C. H. Keitel, V. Mironov, C. J. Osborne, G. Sikler,
R. Soria Orts, V. Shabaev, H. Tawara, I. I. Tupitsyn, J. Ullrich, and A. Volotka,  
Phys. Rev. Lett. {\bf 95}, 183001 (2005).

\bibitem{Brenner2007} 
G. Brenner, J. R. Crespo L\'{o}pez-Urrutia, Z. Harman, P. H. Mokler, and J. Ullrich, 
Phys. Rev. A {\bf 75}, 032504 (2007).

\bibitem{Tupit05}
I. I. Tupitsyn, A. V. Volotka, D. A. Glazov, V. M. Shabaev, G.
Plunien, J. R. Crespo L\'opez-Urrutia, A. Lapierre, and J. Ullrich,
Phys. Rev. A {\bf 72}, 062503 (2005).

\bibitem{volotka06} 
A.V. Volotka, D.A. Glazov, G. Plunien, V.M. Shabaev, and I.I. Tupitsyn,
Eur. Phys. J. D {\bf 38}, 293 (2006).

\bibitem{Volotka08}
A.V. Volotka, D.A. Glazov, G. Plunien, V.M. Shabaev, and I.I. Tupitsyn,
Eur. Phys. J. D {\bf 48}, 167 (2008).

\bibitem{Rynkun12}
P. Rynkun, P. J\"onsson, G. Gaigalas, and C. F. Fischer
At. Data Nucl. Data Tables {\bf 98}, 481 (2012).


\bibitem{Marq12}
J.P. Marques, P. Indelicato, and F. Parente,
Eur. Phys. J. {\bf 66}, 324 (2012).

\bibitem{Fisch16}
C. F. Fischer, I. P. Grant, G. Gaigalas, and P. Rynkun, 
Phys. Rev. A {\bf 93}, 022505 (2016). 

\bibitem{Vilkas03}
M. J. Vilkas and Y. Ishikawa,
Phys. Rev. A {\bf 68}, 012503 (2003).

\bibitem{Sant09}
J. A. Santana, Y. Ishikawa, and E. Tr\"abert,
Phys.Scr. {\bf 79}, 065301 (2009).

\bibitem{Jonson13}
P. J\"onsson, G. Gaigalas, J. Biero\'{n}, C. Froese Fischer and I. P. Grant,
Comput. Phys. Commun., {\bf 184}, 2197 (2013).

\bibitem{Grant07}
I. P. Grant, 
\textit{Relativistic Quantum Theory of Atoms and Molecules},
(Springer, New York, 2007). 

\bibitem{Dyall}K. G. Dyall, I. P. Grant, C. T. Johnson, F. A. Parpia and  E. P. Plummer, 
Comput. Phys. Commun. {\bf 55}, 424 (1989).

\bibitem{volotka06D} A. V. Volotka, High-precision QED calculations of the hyperfine
 structure in hydrogen and transition rates in multicharged ions,
 PhD Thesis, Technische Universit\"at Dresden (2006).

 \bibitem{Nandy16}
D. K. Nandy,
Phys. Rev. A {\bf 94}, 052507 (2016).

\bibitem{Jonnson13}
P. J\"onsson, A. Alkauskas, and G. Gaigalas,
At. Data Nucl. Data Tables {\bf 99} 431 (2013).

\bibitem{Jonaus04}
V. Jonauskas, F. P. Keenan, M. E. Foord, R. F. Heeter, S. J. Rose, P. A. M. van Hoof,
G. J. Ferland, K. M. Aggarwal, R. Kisielius, and P. H. Norrington,
Astron. Astrophys. {\bf 416}, 383 (2004).

\bibitem{Aggar16}
K. M. Aggarwal and F. P. Keenan, 
At. Data Nucl. Data Tables {\bf 109-110} 205 (2016).

\bibitem{Aggar16b}
K. M. Aggarwal and F. P. Keenan, 
At. Data Nucl. Data Tables {\bf 111-112} 187 (2016).

\bibitem{Nandy14}
D. K. Nandy and B. K. Sahoo,
Astron. Astrophys. {\bf 563}, A25 (2014).

\bibitem{Moehs01}
D. P. Moehs, M. I. Bhatti, and D. A. Church,
Phys. Rev. A {\bf 63}, 032515 (2001).

\bibitem{Aggar14}
K. M. Aggarwal and F. P. Keenan,
At. Data Nucl. Data Tables {\bf 100} 1603 (2014).

\bibitem{Ishikawa10}
Y. Ishikawa, J. A. Santana, and E. Tr\"abert,
J. Phys. B {\bf 43}, 074022 (2010).

\bibitem{Gajen16}
G. Singh and N. K. Puri,
J. Phys. B {\bf 49}, 205002 (2016)

\bibitem{Derev12} H. Gharibnejad and A Derevianko,
Phys. Rev. A {\bf 86}, 022505 (2012).

\bibitem{NIST}A. Kramida, Yu. Ralchenko, J. Reader, and NIST ASD Team,
\textit{NIST Atomic Spectra Database} (ver. 5.2.2) (2018),
[Online]. Available: http://physics.nist.gov/asd 

\bibitem{Brenner09}
G. Brenner, J. R. Crespo L\'{o}pez-Urrutia, S. Bernitt, 
D. Fischer, R. Ginzel, K. Kubi\v cek, V. M\"ackel,
P. H. Mokler, M. C. Simon, and J. Ullrich,
Astrophys. J. {\bf 703}, 68 (2009).

\bibitem{Trabert04}
E. Tr\"abert, G. Saathoff and A Wolf,
J. Phys. B {\bf 37}, 945 (2004).

\bibitem{Epp07}S. W. Epp,
Phys. Rev. Lett. {\bf 98}, 183001 (2007).

\bibitem{Trabert14a}E. Tr\"abert,
Appl. Phys. B {\bf 114} 167 (2014).


  





\end{thebibliography}
\end{document}